\documentclass[doublecol]{epl2} 
\usepackage{lineno,hyperref}
\usepackage{amsmath}
\usepackage{amsfonts}
\usepackage{amssymb}
\usepackage{mathrsfs}
\usepackage{mathabx}
\usepackage{booktabs}
\usepackage{bbm}
\usepackage[ruled]{algorithm2e}
\modulolinenumbers[5]
\usepackage{threeparttable}

\title{Predicting hyperlinks via hypernetwork loop structure}

\author{Liming Pan\inst{1} \and Hui-Juan Shang\inst{1} \and Peiyan Li\inst{2} \and Haixing Dai\inst{3} \and Wei Wang\inst{4}\footnote{Correspondence to: wwzqbx@hotmail.com} \and Lixin Tian\inst{5}\footnote{Correspondence to: tianlx@ujs.edu.cn}}
\shortauthor{L. Pan \etal}

\institute{                    
  \inst{1} School of Computer and Electronic Information, Nanjing Normal University - Nanjing 210023, China\\
  \inst{2} Institute of Computer Science, LMU Munich - Munich 80538, Germany\\
  \inst{3} The University of Georgia - Athens, GA 30602 USA.\\
  \inst{4} Cybersecurity Research Institute, Sichuan University - Chengdu 610065, China\\
  \inst{5} School of Mathematical Sciences, Nanjing Normal University - Nanjing 210023, China
}

\abstract{
While links in simple networks describe pairwise interactions between nodes, it is necessary to incorporate hypernetworks for modeling complex systems with arbitrary-sized interactions. In this study, we focus on the hyperlink prediction problem in hypernetworks, for which the current state-of-art methods are latent-feature-based. A practical algorithm via topological features, which can provide understandings of the organizational principles of hypernetworks, is still lacking. For simple networks, local clustering or loop reflects the correlations among nodes; therefore, loop-based link prediction algorithms have achieved accurate performance. Extending the idea to hyperlink prediction faces several challenges. For instance, what is an effective way of defining loops for prediction is not clear yet; besides, directly comparing topological statistics of variable-sized hyperlinks could introduce biases in hyperlink cardinality. In this study, we address the issues and propose a loop-based hyperlink prediction approach. First, we discuss and define the loops in hypernetworks; then, we transfer the loop-features into a hyperlink prediction algorithm via a simple modified logistic regression. Numerical experiments on multiple real-world datasets demonstrate superior performance compared to the state-of-the-art methods.}

\begin{document}

\maketitle

\section{Introduction}
Networks have been a powerful tool for modeling interacting complex systems ranging from social, technological, and biological systems~\cite{albert2002statistical,newman2003structure}. For instance, networks can be adopted to abstract the friendship between pairs of people, the interconnections between routers of the Internet, and the interactions between biological molecules. Due to technical limitations or experimental errors, the network we observed can be incomplete. Link prediction (LP) algorithms~\cite{liben2007link,lu2011link,lu2015toward,zhang2020predicting,pech2017link} aim at finding missing links based on the observed network data. Besides, link prediction algorithms can also forecast future links on time-evolving systems.

Despite the success in modeling a wide variety of systems as networks, recent studies have realized that traditional simple networks have a fundamental limit: they capture only pairwise interactions in the system~\cite{benson2018simplicial,battiston2020networks}. Take the collaborations in a co-authorship network as an example: an article could involve a group of authors rather than two; therefore, describing the co-authorship via pairwise relationship ignores higher-order correlations~\cite{patania2017shape}. Higher-order relations are ubiquitous in real-world systems and data. Other examples include the relationship among the reactants of a chemical reaction\cite{shen2018genome,jost2019hypergraph}, higher-order correlations in a neural population~\cite{schneidman2006weak}, the interference among species in ecology~\cite{bairey2016high}, communications or interactions for people in social groups~\cite{benson2018simplicial}, to name just a few. In order to model these higher-order interactions, hypernetworks and dynamics on hypernetworks have attracted vast attention in recent studies~\cite{de2021phase,alvarez2021evolutionary,reitz2020higher}. 

Like the traditional LP problem, the target of hyperlink prediction (HLP) is to predict missing higher-order relationships in a hypernetwork. Despite the ubiquitousness of hypernetworks, studies on hyperlink prediction are still relatively limited. We can roughly categorize LP methods as topological feature-based, which uses topological statistics, and latent feature-based, which embeds the nodes in a latent space. A state-of-art method is the Coordinated Matrix Minimization (CMM)~\cite{zhang2018beyond}, which employs a latent-space approach. As shown in ref.~\cite{zhang2018beyond}, although many algorithms for the traditional link prediction adopt the topological-based approach, e.g., common neighbors (CN)~\cite{liben2007link}, Adamic–Adar coefficient  (AA)~\cite{adamic2003friends}, and Katz similarity (Katz)~\cite{katz1953new}, they are not directly applicable for HLP.

A practical topological feature-based HLP approach not only can be adopted for applications but also provides insight in understanding the organizational principles of real-world complex systems~\cite{lu2011link}. For HLP, a topological feature-based method is still lacking, as several challenges are to be addressed. Firstly, traditional LP methods evaluate the topological statistics in a local neighborhood of the focal candidate link, which are usually defined only for pairwise nodes, and it is not clear yet how to extend them for higher-order relations. For example, CN assigns a score to each candidate link by the number of length-two walks between its two ends. However, a naive generalization by averaging the CN scores among all node pairs in a hyperlink performs poorly~\cite{zhang2018beyond}. Secondly, the statistics of local topological features depend on the cardinality of the focal hyperlink. A hyperlink involving more nodes should contain more common neighbors if we sum up all node pairs; therefore, we could introduce implicit biases on the hyperlink cardinality for prediction without comparing different-sized hyperlinks on the same ground. In general, we have to overcome the problem of comparing hyperlinks of variable sizes for topological feature-based approaches.

In simple networks, traditional LP algorithms have successfully adopted walks and loops features for prediction. A $\tau$-walk is any sequence of $\tau$ nodes such that every consecutive pair of nodes is connected by an edge~\cite{newman2018networks}. In LP, CN can be interpreted as counting the number of $2$-walks between two nodes, while Katz is a weighted sum over walks of all lengths with the weight decaying exponentially with the length. A walk that starts and ends at the same node is called a loop. Ref.~\cite{pan2016predicting} defines an exponential random graph model in terms of loops and has achieved good prediction accuracy for the LP task. Whether the features of walk or loop of a hypernetwork can be adopted for the HLP problem is not clear yet, as CN and Katz's simple generalizations do not perform well~\cite{zhang2018beyond}.

In this study, we propose a HLP method via loop-features. First, we define two types of loops, namely node-based and hyperlink-based, in hypernetworks. Based on the definitions, we score a candidate hyperlink by how it shapes the loop structure of the underlying hypernetwork. Hyperlinks of different sizes are related and compared via a scaling function. In the end, we turn the loop-features into a HLP algorithm via a simple modified logistic regression. Through numerical experiments, the proposed algorithm demonstrates superior performance compared to the state-of-the-art methods. From the results, we find that it is necessary to consider the usual node-based loops and the dual hyperlink-based loops simultaneously for the HLP task, while for LP, only the former might be sufficient.

\section{Preliminaries}
A hypernetwork is an order pair $G=(V,E)$, where $V=\{v_1,\cdots,v_n\}$ is the set of $n$ nodes and $E=\{e_1,\cdots,e_m\}\subseteq 2^V$ is the set of $m$ hyperlinks. A hyperlink $e_{a}\subset V$ describes a higher-order or group-based relations among nodes. When $\vert e_{a}\vert=2$ for all $a\in [m]$, the hyergraph reduces to a simple graph with pairwise connections. The structure of a hypernetwork can be represented by the incidence matrix $\mathbf{S}\in \{0,1\}^{n\times m}$, which has entries $\mathbf{S}_{i,a}=\mathbb{I}\left(v_i\in e_{a}\right)$ with $\mathbb{I}\left(\cdot\right)$ the indicator function.

When the hypernetwork is incomplete and only observes part of the hyperlinks $E^c\subset E$, HLP aims to reveal missing hyperlinks from $\widebar{E}=2^V-E^o$. For a traditional link prediction problem, i.e., when $\vert e_{a}\vert=2$ for all $a\in [m]$, there are ${n\choose 2} - m$ candidate edges; whereas for hyperlink prediction, the cardinality of all hyperlinks is $\vert \widebar{E} \vert= 2^n-m$. Fortunately, as suggested in ref.~\cite{zhang2018beyond}, in many cases, we can filter out irrelevant hyperlinks and focus on a subset of as $E^c\subset \widebar{E}$. For instance, in biological or chemical reactions hypernetworks, most hyperlinks have no meaning; meanwhile, it is rare for the coauthorship hypernetwork to find papers with more than ten authors for many research areas.

\section{Walks and loops in hypernetworks}
For a simple network, we define a walk of length $\tau$ as a sequence of nodes $W=(v_{i_1},v_{i_2},\cdots ,v_{i_{\tau+1}})$ such that every consecutive nodes is connected by an edge. The walk $W$ is called a loop whenever $v_{i_1}=v_{i_{\tau+1}}$, i.e., when the walk starts and ends at the same node. The way of generalizing walks in hypernetworks is not unique, and there have been several alternatives~\cite{battiston2020networks}. For instance, a $k$-walk of length $\tau$ is defined as a sequence of hyperlinks such that $\vert e_{a}\cap e_{a+1}\vert=k$ with $e_a\neq e_{a+1}$ for all $a\in [\tau-1]$. Alternatively, we can define a $k$-walk as a sequence of hyperlinks such that each pair of consecutive hyperlinks intersect in at least $k$ nodes~\cite{aksoy2020hypernetwork}. Besides, ref.~\cite{carletti2021random} designs random-walks by introducing a weight to each hyperlink according to its cardinality.

\begin{figure} [!htbp]
\centering
\includegraphics[width=0.9\linewidth]{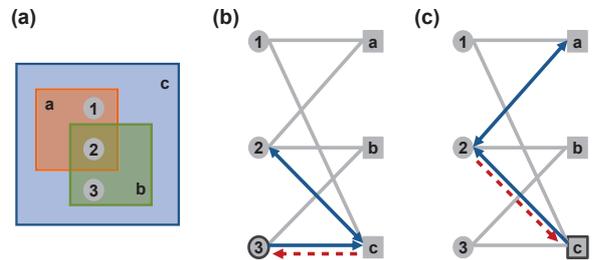}
\caption{An illustration of walks in hypernetworks. (a) A hypernetwork with nodes $\{1,2,3\}$ and hyperlinks $\{a,b,c\}$. (b) Node-based walks and (c) hyperlink-based walks which are non-backtracking in nodes and hyperlinks, respectively.}
\label{fig1}
\end{figure}

In the following, we propose a generalization of walks in hypernetworks for the LP task. Consider the walks in simple graphs and we describe it at the following two-step process. (\romannumeral1) Staring from node $v_{i_1}$,  pick any hyperlink such that $v_{i_1}\in e_{a_1}$. (\romannumeral2) Move along the hyperlink to any node in $e_{a_1}$ other than $v_{i_1}$. Then we repeated the procedure for $\tau$ times to obtain a walk of length $\tau$. As a consequence, a $\tau$-walk is defined as a sequence of alternating nodes and hyperlinks $(v_{i_1},e_{a_1},v_{i_2},e_{a_2},\cdots e_{a_{\tau}},v_{i_{\tau+1}})$. For simple networks, the destination node in step (\romannumeral2) is unique, and we can omit specifying the links in the walk and denote it as $(v_{i_1},v_{i_2},\cdots ,v_{i_{\tau+1}})$, which reduces to the traditional definition of walks.

With the above definition, we can conveniently count the number of walks between any pair of nodes via algebraic manipulations of the incidence matrix $\mathbf{S}$. Defined the adjacency matrix as 
\begin{equation}
\mathbf{A}=\mathbf{S}\mathbf{S}^\intercal-\mathbf{D},
\end{equation}
where $\mathbf{D}$ is the diagonal matrix whose diagonal entries are the number of hyperlinks that a node belongs to. For any two nodes $i,j$, the number of ways moving from $i$ to $j$ equals the number of hyperlinks that both the two nodes belong, which is $(\mathbf{S}\mathbf{S}^\intercal)_{ij}$. As we forbid the walk to stay at the same node in each step, the diagonal entries of $\mathbf{S}\mathbf{S}^\intercal$ are set to zero by subtracting $\mathbf{D}$. Therefore, the total number of $\tau$-walks between node $i$ and $j$ is $(\mathbf{A}^{\tau})_{ij}$.

For a simple graph, the incidence matrix $\mathbf{S}$ can be recovered from the adjacency matrix $\mathbf{A}$ up to a relabeling of the links; therefore the information of structures has not been reduced upon projecting into the nodes. However, in general this is not true for hypernetworks. For instance, consider two hypernetworks defined on the same set of nodes $V=\{v_1,v_2,v_3\}$, but with edge sets $E=\{\{v_1,v_2\},\{v_1,v_3\},\{v_2,v_3\}\}$ and $E=\{\{v_1,v_2,v_3\}\}$, respectively. It can be checked that the two hypernetworks result in the same adjacency matrix $\mathbf{A}$. Therefore, to complement the node-based walks, we introduce a dual hyperlink-based walks. Recall that in the definition of a walk, i.e., $(v_{i_1},e_{a_1},v_{i_2},e_{a_2},\cdots e_{a_{\tau}},v_{i_{\tau+1}})$, we have ensured a walker cannot staying at the same node, i.e., $v_{i_n}\neq v_{i_{n+1}}$, along the walk. Intuitively, the definition reflects how nodes are related to its neighbors. To characterize the correlations among hyperlinks, we define the dual concept of walks that starts from a hyperlink without repeating consecutive hyperlinks, i.e., $e_{a_n}\neq e_{a_{n+1}}$. Define the intersection profile~\cite{battiston2020networks} as 
\begin{equation}
\mathbf{P}=\mathbf{S}^\intercal\mathbf{S}-\mathbf{Z},
\end{equation}
where $\mathbf{Z}$ is the diagonal matrix whose diagonal entries are the cardinlities of hyperlinks. The number of length $\tau$ hyperlink-based walks between hyperlinks $a$ and $b$ is $(\mathbf{P}^{\tau})_{a b}$. 

In summary, we have defined two types of walks in hypernetworks, namely node-based and hyperlink-based, which are non-backtracking in nodes and hyperlinks, respectively. Fig.~\ref{fig1} illustrates the two types of walks. In Fig.~\ref{fig1}(a), we show a hypernetwork with three nodes $V=\{1,2,3\}$ and three hyperlinks $E=\{a,b,c\}$, where the gray circles represent nodes and the nodes are connected by a hyperlink if they lie inside the same square. Consider node-based walks starting from node $3$ in the illustrated hypernetwork. First, we pick any hyperlink, say $c$, which contains $3$. In the next step, we have to pick any node in $c$ other than $3$. Thus
the sequence $(3,c,3)$ is not a feasible walk as shown by the red dashed arrow in Fig.~\ref{fig1}(b). Meanwhile, $(3,c,2,c)$ is a well-defined node-based walk as it only backtracks in hyperlinks. Similarly, hyperlink-based walks do not allow backtracking in hyperlinks. Suppose we start with the hyperlink $c$ and move to node $2$; then, in the next step, we cannot move back to $c$ immediately, as shown in Fig.~\ref{fig1}(c). Meanwhile, $(c,2,a,2)$ is a feasible hyperlink-based walk. Note that even two hypernetworks have the same $\mathbf{A}$ and $\mathbf{P}$, they are not necessarily isomorphic, as counter-examples are shown in ref.~\cite{kirkland2018two}.

Similar to simple graphs, a loop is a walk that starts and ends at the same node. The total number of node-based and hyperlink-based $\tau$-loops are therefore $\mathrm{tr}(\mathbf{A}^{\tau})$ and $\mathrm{tr}(\mathbf{P}^{\tau})$, respectively, where $\mathrm{tr}(\cdot)$ is the matrix trace.

\begin{table*}[ht!]
\centering
 \begin{tabular}{ccccccccc}
\toprule
Dataset & iJO1366 & iAF1260b & iAF692 & iHN637 & iIT341 & iAB\_RBC\_283 & Enron-email & NDC-classes\\
\midrule
n & 1805 & 1668 & 628 & 698 & 485 & 342 & 148 & 1161 \\
m & 2583 & 2388 & 690 & 785 & 554 & 469 & 1512 & 1088 \\
\bottomrule
\end{tabular}
\caption{Number of nodes and hyperlinks for the eight datasets.\label{table1}}
\end{table*}

\section{Hyperlink prediction via loops}
In this section, we transfer the loop features into a HLP algorithm with the following steps. First, we estimate how the spectrum of loops is altered by adding a candidate hyperlink to the hypernetwork. Then we define a score function for each hyperlink as a weighted sum of these changes. Finally, we make predictions via a modified logistic regression with the score function as the predictor. 

For any hyperlink $e$, let $G_{e+}=(V,E\cup \{e\})$ and $G_{e-}=(V,E\setminus \{e\})$ to be the hypernetwork that the hyperlink $e$ is forced to be present or absent from $G$, respectively. For a hypernetwork $G$, define the following function of as a weighted sum over loops with different length 
\begin{equation}\label{eq:Hamiltonian}
S(G)=\sum_{\tau=2}^{\tau_c}\alpha_{\tau} \log \mathrm{tr} \left( \mathbf{A}^{\tau}\right)+\sum_{\tau=2}^{\tau_c}\beta_{\tau} \log \mathrm{tr} \left(\mathbf{P}^{\tau}\right),
\end{equation}
where $\{\alpha_{\tau},\beta_{\tau}\}$ are the weight parameters and $\tau_c$ is the cutoff of the loop length.
The justification of the definition is as follows. Consider the matrix $\mathbf{A}$ and let $\{\omega_i:i\in [n]\}$ be its eigenvalues, then we have $\mathrm{tr}(\mathbf{A}^{\tau})=\sum_{i=1}^n \omega^{\tau}_i=\omega^{\tau}_1\sum_{i=1}^n (\omega_i/\omega_1)^{\tau}$. Therefore, when $\tau$ is large we have $\mathrm{tr}(\mathbf{A}^{\tau})\approx \omega_1^{\tau}$. In other words, the number of loops grows exponentially with $\tau$ and we take the logarithm of $\mathrm{tr} (\mathbf{A}^{\tau})$ in eq.~\eqref{eq:Hamiltonian} to make each term in the summation in the same order of magnitude. Besides, as $\mathrm{tr}(\mathbf{A}^{\tau+1}) \approx \omega_1 \mathrm{tr}(\mathbf{A}^{\tau})$ for large $\tau$, including longer loops do not introduce further information of the hypegraph structures and causes multicollinearity in the explanatory variables; thus we introduce the length cutoff $\tau_c$. 

For any $e\in E^o\cup E^c$, the difference $S\left(G_{e+}\right)-S\left(G_{e-}\right)$ quantifies how much the loop structure is altered by the presence of $e$. Let $\mathbb{P}(e\in E)$ be the probability that the hyperlink $e$ is a true hyperlink, either observed or missing. To make predictions, we assume its log-odds is given by
\begin{equation}\label{eq:log_odd}
\log \frac{\mathbb{P}(e\in E)}{1-\mathbb{P}(e\in E)} = c+ \frac{1}{\vert e\vert^{\gamma}} \left[ S\left(G_{e+}\right)-S\left(G_{e-}\right)\right],
\end{equation} 
where $\gamma$ and $c$ are parameters to be determined. When $\gamma$ is fixed, eq.~\eqref{eq:log_odd} is a standard logistic regression model with parameters $\{\alpha_{\tau},\beta_{\tau}\}$. The scaling function $\vert e \vert^{-\gamma}$ is introduced to relate and compare hyperlinks of different sizes. As discussed above, the term $S\left(G_{e+}\right)-S\left(G_{e-}\right)$ in eq.~\eqref{eq:log_odd} quantifies how a hyperlink $e$ shapes the loop structure of a hypernetwork. However, for a hyperlink with larger cardinality, $\vert e\vert$ should in general change the structure of the hypernetwork in a more dramatic way intuitively; therefore, we cannot compare $S\left(G_{e+}\right)-S\left(G_{e-}\right)$ for hyperlinks with different cardinality directly. To address the arbitrary-sized hyperlink cardinality problem, we introduce the heuristic scaling function $\vert e\vert^{-\gamma}$. The scaling function decreases with $\vert e\vert$ for $\gamma>0$, thus punishes larger hyperlinks. We will show that with the heuristic scaling function, the proposed method performs well; nevertheless, a better way of relating different-sized hyperlinks requires further discussions.

To optimize the model, we label $L(e)=+1$ for $e\in E^o$ as positive examples and $L(e)=-1$ for $e\in E^c$ as negative ones. Then, for fixed $\gamma$, we obtain $\{\alpha_{\tau},\beta_{\tau}\}$ by maximizing the following likelihood function
\begin{equation}
\begin{split}
\mathcal{L}\left(\{\alpha_{\tau},\beta_{\tau}\}\vert\gamma \right)= \prod_{e\in E^o\cup E^c} &\left[\mathbb{P}(e\in E)\right]^{\mathbb{I}(e\in E^o)}\\
\times &\left[1-\mathbb{P}(e\in E)\right]^{(1-\mathbb{I}(e\in E^o))},
\end{split}
\end{equation}
where $\mathbb{P}(e\in E)$ is defined by eq.~\ref{eq:log_odd}. We can solve the maximization can by Gauss-Newton method or any logistic regression model package. 
 The optimal set of parameters 
$\{\alpha^*_{\tau},\beta^*_{\tau},\gamma^*\}$ is determined by a line search over $\gamma$. Concretely, for each $\gamma$ from $0$ to $2$ with a step of $0.1$, we find the corresponding optimal $\{\alpha_{\tau},\beta_{\tau}\}$. Then we compare different choices of $\gamma$ to find the overall optimal parameters. One remaining parameter to be determined is the length cutoff $\tau_c$, and we take it as a hyperparameter.

\section{Experiments}

\begin{table*}[ht!]
 \begin{tabular}{ccccccccccccc}
\toprule
Dataset & Ours & CMM & BS & SHC & Katz & CN\\
\midrule
iJO1366 & $0.511\pm 0.009$ & $\mathbf{0.522}\pm \mathbf{0.025}$ & $0.438\pm 0.010$ & $0.420\pm 0.010$ & $0.296\pm 0.010 $ & $0.192\pm 0.006$\\
iAF1260b & $\mathbf{0.556}\pm \mathbf{0.006}$ & $0.474\pm 0.018$ & $0.415\pm 0.010$ & $0.442\pm 0.008$ & $0.289\pm 0.038$ & $0.220\pm 0.003$\\
iAF692 & $\mathbf{0.530}\pm \mathbf{0.019}$ & $0.457\pm 0.027$ & $0.363\pm 0.028$ & $0.373\pm 0.011$ & $0.237\pm 0.047$ & $0.210\pm 0.013$\\
iHN637 & $\mathbf{0.452}\pm \mathbf{0.024}$ & $0.434\pm 0.036$ & $0.290\pm 0.024$ & $0.315\pm 0.021$ & $0.269\pm 0.067$ & $0.155\pm 0.023$\\
iIT341 & $\mathbf{0.463}\pm \mathbf{0.022}$ & $0.429\pm 0.020$ & $0.268\pm 0.021$ & $0.333\pm 0.010$ & $0.141\pm 0.050$ & $0.196\pm 0.013$\\
iAB\_RBC\_283 & $\mathbf{0.630}\pm \mathbf{0.020}$ & $0.528\pm 0.052$ & $0.504\pm 0.020$ & $0.560\pm 0.019$ & $0.275\pm 0.023$ & $0.339\pm 0.009$\\
Enron-email & $\mathbf{0.710}\pm \mathbf{0.024}$ & $0.367\pm 0.008$ & $0.243\pm 0.010$ & $0.278\pm 0.010$ & $0.595\pm 0.020$ & $0.476\pm 0.011$\\
NDC-classes & $\mathbf{0.843}\pm \mathbf{0.013}$ & $0.390\pm 0.005$ & $0.198\pm 0.010$ & $0.233\pm 0.017$ & $0.645\pm 0.032$ & $
0.347\pm 0.017$\\
\bottomrule
\end{tabular}
\caption{The prediction accuracy measure by Precision for the eight datasets.\label{table2}}
\end{table*}

\begin{table*}[ht!]
\centering
 \begin{tabular}{ccccccccccccc}
\toprule
Dataset & Ours & CMM & BS & SHC & Katz & CN\\
\midrule
iJO1366 & $\mathbf{0.724}\pm \mathbf{0.002}$ & $0.709\pm 0.018$ & $0.682\pm 0.010$ & $0.711\pm 0.004$ & $0.512\pm 0.008 $ & $0.437\pm 0.010$\\
iAF1260b & $\mathbf{0.739}\pm \mathbf{0.009}$ & $0.702\pm 0.003$ & $0.670\pm 0.013$ & $0.715\pm 0.005 $ & $0.535\pm 0.008$ & $0.468\pm 0.003$\\
iAF692 & $\mathbf{0.741}\pm \mathbf{0.011}$ & $0.704\pm 0.026$ & $0.506\pm 0.029$ & $0.616\pm 0.018 $ & $0.506\pm 0.020$ & $0.430\pm 0.021$\\
iHN637 & $\mathbf{0.734}\pm \mathbf{0.012}$ & $0.705\pm 0.033$ & $0.526\pm 0.027$ & $0.617\pm 0.014 $ & $0.531\pm 0.021$ & $0.424\pm 0.021$\\
iIT341 & $\mathbf{0.718}\pm \mathbf{0.017}$ & $0.679\pm 0.015$ & $0.511\pm 0.023$ & $0.598\pm 0.012 $ & $0.496\pm 0.018$ & $0.440\pm 0.010$\\
iAB\_RBC\_283 & $\mathbf{0.791}\pm \mathbf{0.018}$ & $0.710\pm 0.048$ & $0.609\pm 0.014$ & $0.696\pm 0.012 $ & $0.512\pm 0.022$ & $0.388\pm 0.012$\\
Enron-email & $\mathbf{0.882}\pm \mathbf{0.004}$ & $0.586\pm 0.007$ & $0.467\pm 0.013$ & $0.523\pm 0.009 $ & $0.791\pm 0.027$ & $0.719\pm 0.006$\\
NDC-classes & $\mathbf{0.966}\pm \mathbf{0.003}$ & $0.574\pm 0.017$ & $0.361\pm 0.009$ & $0.375\pm 0.009 $ & $0.760\pm 0.013$ & $0.526\pm 0.009$\\
\bottomrule
\end{tabular}
\caption{The prediction accuracy measure by AUC for the eight datasets.\label{table3}}
\end{table*}

In this section, we evaluate the proposed method on real-world hypernetwork datasets.

\subsection{Datasets} We conduct experiments on eight real-world hypernetworks, including metabolic reactions, email communications, and the drug-classes labels. The metabolic hypernetwork takes metabolites as nodes and reactions among metabolites as hyperlinks. The problem of predicting vial missing hyperlinks was considered in ref.~\cite{zhang2018beyond}. We conduct experiments on six metabolic hypernetwork in ref.~\cite{zhang2018beyond}, which includes (1) iJO1366, (2) iAF1260b, (3) iAF692, (4) iHN637, (5) iIT341 and (6) iAB\_RBC\_283,  and use their candidate set $E^c$. The number of nodes and hyperlinks for the six metabolic hypernetworks can be found in table~\ref{table1}. For the two largest datasets, i.e., (1) iJO1366 and (2) iAF1260b, we randomly delete $400$ reactions as missing hyperlinks and the remaining ones as the observed hyperlinks. For the rest four datasets, we set the size of the test set as $200$. More details of constructing the metabolic reaction hypernetworks can be found in ref.~\cite{zhang2018beyond}.

The email network (Enron-email) contains emails generated by employees of the Enron Corporation~\cite{benson2018simplicial,klimt2004enron}. We build the hypernetwork where nodes are email addresses, and a hyperlink includes all recipient addresses on an email and the sender's address. We ignore each hyperlink's time stamps and remove all repeated hyperlinks to focus on the static structure. The number of nodes and hyperlinks of the resulting hypernetwork is shown in table~\ref{table1}. For the experiments, we randomly delete $400$ hyperlinks as missing ones and generate $1200$ fake hyperlinks according to the hyperlink distribution and nodal degree distribution. Concretely, we generate a random integer number according to the hyperlink cardinality distribution; then, we pick the generated number of nodes with a probability proportional to the nodal degrees. 

The drug network (NDC-classes) represents the class labels of drugs from the National Drug Code Directory~\cite{benson2018simplicial}. The nodes are class labels, and a hyperlink is the class labels for a drug. The statistics of NDC-classes are shown in table~\ref{table1}. To conduct experiments, we randomly delete $400$ hyperlinks as the missing ones and generate $1200$ fake hyperlinks according to the hyperlink distribution and nodal degree distribution.

\subsection{Baselines and experimental setting}
We compare the proposed methods with several previous approaches, including simple generalizations of topological feature-based methods and state-of-the-art methods. For topological feature-based methods, we consider the generalized CN and Katz similarity. For the two methods, we compute the score for each pair of nodes as simple graphs using the adjacency matrix $\mathbf{A}$, then the average gives the score for a hyperlink among all the nodes pairs it contains. For Katz, we choose the parameter (i.e., the damping factor) via cross-validation. Besides, we also consider the Bayesian Set (BS)~\cite{ghahramani2005bayesian} method, an information retrieval algorithm that retrieves similar hyperlinks of $E^c$ from $E^o$.

Two state-of-the-art methods are Spectral hypernetwork Clustering (SHC)~\cite{zhou2006learning} and CMM~\cite{zhang2018beyond}. SHC is the generalized version of the simple graph spectral clustering technique and has been adopted for HLP in ref.~\cite{zhang2018beyond}. CMM performs nonnegative matrix factorization and least square matching in the vertex adjacency space of the hypernetwork alternately to infer a subset of candidate hyperlinks that are most suitable to fill the training hypernetwork. Hyperparamters in CMM are determined the same way as in ref.~\cite{zhang2018beyond}.

For our proposed approach, it contains one hyperparameter, which is the length cutoff $\tau_c$. For two relatively large datasets with more nodes and hyperlinks, i.e.,(1) iJO1366, (2) iAF1260b, we set $\tau_c$ to the default $8$.
For the rest datasets, we pick $\tau_c$ in $\{6,7,\cdots,14\}$ via cross-validation on the training set.

\subsection{Results}

\begin{figure} [!htbp]
\centering
\includegraphics[width=0.9\linewidth]{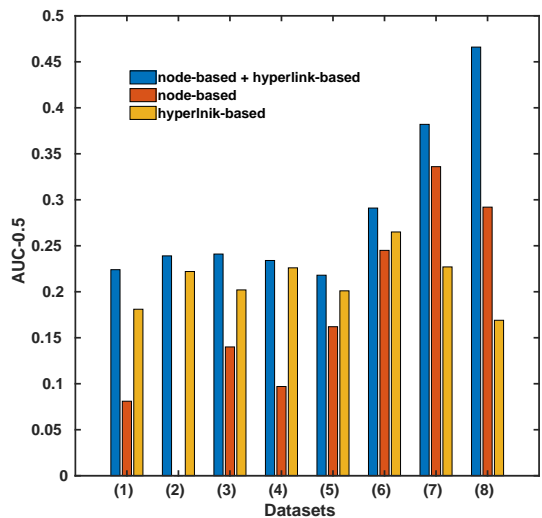}
\caption{AUC of predictions with (a) nodes-based plus hyperlink based loops, (b) only node-based loops, and (c) only hyperlink-based loops. We have subtracted the AUC by 0.5 for better visualization.}
\label{fig2}
\end{figure}

We measure the prediction accuracy by two evaluation metrics, namely area under the ROC curve (AUC) and Precision~\cite{lu2011link}. We perform 12 independent experiments for each dataset. The average and standard deviation of Precision is shown in table~\ref{table2} and of AUC in table~\ref{table3}. The proposed approach outperforms other baselines except for the iJO1366 dataset when measured by Precision. Especially for the Enron-email and NDC-classes dataset, the loop-based approach predicts the missing hyperlinks in good accuracy when other approaches do not work well. The experimental results indicate that the loop-features can successfully capture the organizational principles of hypernetworks.

In the experiments, we have incorporated both node-based and hyperlink-based loops, while for LP on simple networks, usually only the former is being considered. As discusses above, for a simple graph, the incidence matrix $\mathbf{S}$ can be recovered from the adjacency matrix $\mathbf{A}$ up to a relabeling of the links, while this is in general not true for hypernetworks. We show by experiments that introducing the complementary hyperlink-based walks does improve the HLP accuracy. We conduct experiments by separately considering only node-based or hyperlink-based loops for prediction. Concretely, we define $S(G)$ in eq.~\ref{eq:Hamiltonian} with only the terms in the first or the second summation and conduct experiments. The predicting accuracy measured by AUC is shown in fig.~\ref{fig2}, where we have subtracted AUC by 0.5 for visualization. Fig.~\ref{fig2} shows that with solely node-based or hyperlink-based loops, the AUC is significantly lower. The results suggest that it is not sufficient to consider the correlations among nodes when characterizing the structures of hypernetworks.

Another phenomenon worth noticing is that although Katz performs poorly for metabolic reaction datasets, it still works for Enron-email and NDC-classes. Therefore, we might conjecture that pairwise relations approximate the higher-order interactions better than other datasets for these two datasets. What is a better way of characterizing higher-order interactions requires further discussions. 

\section{Conclusions}
In this study, we propose a loop-based hyperlink prediction approach. First, we discuss the intuition of walks and loops in simple graphs, and then we generalize the concept to hypernetworks. We have defined two types of loops, namely, node-based and hyperlink-based. We evaluate the tendency to observe a hyperlink by taking it as a perturbation to the observed network and check how the perturbation shapes the loop spectrum of the underlying hypernetwork. Hyperlinks of different sizes are related and compared via a scaling function of its cardinality. Then take a weighted sum of changes in loop spectrum as the predictor for a HLP algorithm. Via simple logistic regressions, we find that the proposed method outperforms the state-of-the-art approaches on the datasets under consideration.

A shortcoming of the proposed method is that it is not computational efficient when applying to large hypernetworks. For each hyperlink in $E^c\cup E^o$, we compute the number of loops before and after flipping it based on $G=(V,E^o)$. For each $e$, we perform the matrix multiplication of $\mathbf{A}$ and $\mathbf{P}$ for $\tau_c-1$ times. Let $\hat{m}=\vert E^c\cup E^o\vert$, therefore, the overall time complexity of the proposed algorithm is $O(\hat{m}(n^3+m^3))$. A solution for solving the problem is to predict the missing hyperlinks in a sub-hypernetwork rather than the entire one. As we only consider loops of finite length, the hyper-loops are localized in the focal hyperlink neighborhood. 

From the experimental results, we find that it is necessary to consider two complementary concepts of loops, i.e., node-based and hyperlink-based, for prediction in hypernetworks. Intuitively, for a simple network, the incidence matrix for a simple graph can be recovered from the adjacency matrix; thereby, it would be relatively sufficient to characterize simple graph organization principles via only node-based loops. Meanwhile, for hypernetworks, we have to track both how the nodes and hyperlinks are organized simultaneously. The phenomenon might provide insight for characterizing and understanding the hyperlink structures and further designing topological feature-based HLP algorithms for hypernetworks.

\acknowledgments
This work is supported by National Natural Science Foundation of China (62006122, 61903266), The Major Program of the National Natural Science Foundation of China (71690242) and the National Key Research and Development Program of China (2020YFA0608601). 


\end{document}